\documentclass{jetpl}
\twocolumn
\lat
\abstract{
Complex fluids, such as polymer solutions and blends, colloids and gels, are
of growing interest in fundamental and applied soft-condensed-matter science.
A common feature of all such systems is the presence of a mesoscopic
structural length scale intermediate between atomic and macroscopic scales.
This mesoscopic structure of complex fluids is often fragile and sensitive to
external perturbations. Complex fluids are frequently viscoelastic (showing a
combination of viscous and elastic behaviour) with their dynamic response
depending on the time and length scales. Recently, non-invasive methods to
infer the rheological response of complex fluids have gained popularity
through the technique of microrheology, where the diffusion of probe spheres
in a viscoelastic fluid is monitored with the aid of light scattering or
microscopy. Here we propose an alternative to traditional microrheology
that does not require doping of probe particles in the fluid (which can
sometimes drastically alter the molecular environment). Instead, our proposed 
method makes use of the phenomenon of "avoided crossing" between modes
associated with the structural relaxation and critical fluctuations that are 
spontaneously generated in the system.}

\title{Probing structural relaxation in complex fluids by critical fluctuations}

\rtitle{Probing structural relaxation in complex fluids by critical fluctuations}

\sodtitle{Probing structural relaxation in complex fluids by critical fluctuations}

\author{A.\,F.\,Kostko$^{*+}$, M.\,A.\,Anisimov$^{*}$\/ \thanks{e-mail:
anisimov@ipst.umd.edu}, and J.\,V.\,Sengers$^{*}$}

\rauthor{A.\,F.\,Kostko$^{+*}$, M.\,A.\,Anisimov$^{*}$\/ \thanks{e-mail:
anisimov@ipst.umd.edu}, and J.\,V.\,Sengers$^{*}$}

\sodauthor{A.\,F.\,Kostko$^{+*}$, M.\,A.\,Anisimov$^{*}$\/ \thanks{e-mail:
anisimov@ipst.umd.edu}, and J.\,V.\,Sengers$^{*}$}

\dates{\today}{*}

\address
{$^*$Institute for Physical Science and Technology and Department of Chemical
Engineering, University of Maryland, College Park, MD 20742, USA\\
~\\
$^+$Department of Physics, St.~Petersburg State University of Refrigeration
and Food Engineering, 9 Lomonosov Str., St.~Petersburg, 191002, Russia}

\begin{document}
\maketitle

A liquid mixture in the vicinity of a critical point of mixing exhibits large
concentration fluctuations. The dynamics of such critical concentration
fluctuations in molecular liquids is well understood: the fluctuations decay
exponentially with a diffusive relaxation time~\cite{Swinney73}%

\begin{equation}
\tau_{q}=\frac{1}{D(q,\xi)q^{2}}\text{ ,}%
\end{equation}

\noindent where $q$ is the wave number of the critical fluctuations, $\xi$ is
the spatial correlation length of the fluctuations and $D$ is the mesoscopic
($q$-dependent) diffusion coefficient. $D$ vanishes at the critical point in
the limit of zero wave number approximately as $\xi^{-1}\sim\varepsilon
^{0.63}$, where $\varepsilon=(T-T_{\mathrm{c}})/T$, the reduced distance
between the temperature $T$ and the critical temperature $T_{\mathrm{c}}$. In
molecular fluids, the $q$-dependent diffusion coefficient is well described by
the mode-coupling theory of critical dynamics
\cite{Kawasaki76,Burstin83}. 

\begin{equation}
D(q,\xi)=\frac{k_{\mathrm{B}}T}{6\pi\xi\eta_{\mathrm{app}}}\frac{K(q,\xi
)}{q^{2}}\left[  1+\left(  \frac{q\xi}{2}\right)  ^{2}\right]  ^{\frac
{z_{\eta}}{2}}\Omega\left(  \frac{\xi}{\xi_{\mathrm{D}}}\right)\text{
,}\label{Eq2Diff}%
\end{equation}

\noindent where $k_{\mathrm{B}}$ is Boltzmann's constant and where the
apparent viscosity $\eta_{\mathrm{app}}$ is expected to be equal to the
solution viscosity $\eta$ measured by macrorheology; $z_{\eta}\simeq0.065$
\cite{Nieuwoudt89} is a universal dynamic scaling exponent. 
The function $K(q,\xi)$ is a universal
(Kawasaki) function with $K(q\xi\rightarrow0)=1$; the function $\Omega\left(
\xi/\xi_{\mathrm{D}}\right)  =2/\pi\arctan(\xi/\xi_{\mathrm{D}})$ is an 
approximated dynamic crossover function, where
$\xi_{\mathrm{D}}$ is a cutoff length for the critical fluctuations
\cite{Kiselev94,Luettmer95}, 

However, new phenomena emerge in a complex fluid where 
$\xi_{\mathrm{D}}$ is a mesoscopic length that may compete with the 
correlation length $\xi$
of the concentration fluctuations. The presence of two mesoscale lengths in
near-critical complex fluids causes the appearance of two dynamic modes: one
will be a diffusive decay of the critical concentration fluctuations and
another one will be a structural relaxation mode, which often reveals itself
as viscoelastic relaxation. The decay time of the diffusive mode can be tuned
over a broad range of time scales by varying the reduced temperature
$\varepsilon$ , so that it may intersect the structural relaxation time, which
is insensitive to the proximity to the critical point. As a consequence it
becomes possible to probe structural relaxation in complex fluids by dynamic
light scattering of critical fluctuations. This method is an alternative to the 
traditional microrheology~\cite{Mason95,Gisler98,Crocker00} that requires 
doping of probe particles in the fluid. 

As an illustration of this principle, we have performed accurate
light-scattering measurements of solutions of nearly monodisperse polystyrene
(with molecular weights $M$ ranging from $10^{4}$ to $10^{7}$) in cyclohexane
\cite{Anisimov02,Kostko02}. The major result of our study is that the critical
dynamics in polymer solutions appears to be very different from that in
molecular fluids. Even for a modest polystyrene molecular weight of $195,900$,
with the dynamic correlation function obeying a single-exponential decay, the
apparent viscosity $\eta_{\mathrm{app}}$ extracted from dynamic light
scattering (DLS) (on the basis of Eq. (\ref{Eq2Diff}) with the correlation
length $\xi$ determined by static light scattering~\cite{Anisimov02}) is
vastly different from both the macroscopic viscosity of the solution and the
viscosity of the solvent (Fig.1). But in terms of this apparent
(''mesoscopic'') viscosity, the mesoscopic diffusion coefficient at various
angles can be well described by Eq. (\ref{Eq2Diff}) (lines through the symbols,
Fig.2). The pertinent question is: what is the physical meaning of this
''mesoscopic'' viscosity determined with DLS? Disagreements between the
predictions of the mode-coupling theory for molecular fluids and the DLS data 
in near-critical polymer solutions have also been noted by 
others~\cite{Lao75,Fine95}, but have not yet been explained.

We have observed an even more dramatic change in dynamics in high
molecular-weight ($M=10^{6}$ and higher) polymer solutions, where the dynamic
correlation function turns out to deviate from a single-exponential decay and
where two dynamic modes are clearly present. Far from the critical point, a
fast mode dominates, while close to the critical point a slow mode dominates.
Between these extremes, the data can be approximated by a sum of two
exponentials, indicating contributions from both modes. The presence of two
dynamic modes near the critical temperature appears to be a universal feature
in macromolecular systems and has been observed also by Ritzl \textit{et
al}.~\cite{Ritzl99} for an $M=1$ million polystyrene solution in cyclohexane
and more recently by Tanaka \textit{et al}.~\cite{Tanaka02} for an $M=4$
million polystyrene solution in diethyl malonate. These modes are effective
dynamic modes, neither of which is purely viscoelastic (dictated by polymer
chain dynamics) or purely diffusive (associated with the decay of critical
fluctuations). Instead, the two observed modes emerge from a coupling of
diffusive and viscoelastic modes, which belong to two different dynamic
universality classes, pertaining to conserved and non-conserved order
parameters~\cite{Halperin77}. The challenge is to quantitatively understand
this coupled dynamic crossover behaviour. A starting point in explaining the
dynamics is the Brochard-De Gennes theory
\cite{Brochard77,Brochard83,BrochardDeGennes83}, which predicts a coupling of
diffusion and chain relaxation in polymer solutions that has been subsequently
detected experimentally in non-critical polymer solutions
\cite{Adam85,Jian96,Nicolai90}.

We submit that the Brochard-De Gennes theory can be applied
to any system with dynamic coupling between conserved and non-conserved order
parameters. Phenomenologically, it follows from this theory that the
time-dependent intensity correlation function is the sum of two exponentials:%

\begin{equation}
g_{2}(t)=1+\left\{  f_{+}\exp\left[  -\frac{t}{\tau_{+}}\right]  +f_{-}%
\exp\left[  -\frac{t}{\tau_{-}}\right]  \right\}  ^{2}\label{Eq3g2}%
\end{equation}

\noindent with two decay times (slow $\tau_{-}$ and fast $\tau_{+}$) and
with corresponding amplitudes ($f\pm$) given by:%

\begin{align}
\frac{1}{\tau_{\pm}} &  =\frac{1+q^{2}\xi_{\mathrm{ve}}^{2}+\frac
{\tau_{\mathrm{ve}}}{\tau_{q}}\pm\sqrt{\left(  1+q^{2}\xi_{\mathrm{ve}}%
^{2}+\frac{\tau_{\mathrm{ve}}}{\tau_{q}}\right)  ^{2}-4\frac{\tau
_{\mathrm{ve}}}{\tau_{q}}}}{2\tau_{\mathrm{ve}}}\text{ ,}\label{Eq4tauf}\\
f_{\pm} &  =\frac{\frac{\tau_{\mathrm{ve}}}{\tau_{\pm}}-\left(  1+q^{2}%
\xi_{\mathrm{ve}}^{2}\right)  }{\frac{\tau_{\mathrm{ve}}}{\tau_{+}}-\frac
{\tau_{\mathrm{ve}}}{\tau_{-}}}\text{ .}\label{Eq5f}%
\end{align}

In Eq. (\ref{Eq4tauf}) $\tau_{\mathrm{ve}}$ is the $q$-independent viscoelastic
relaxation time, $\tau_{q}$ is the $q$-dependent diffusion relaxation time,
and $\xi_{\mathrm{ve}}$ the mesoscopic viscoelastic length~ \cite{Doi92}. The
above theory indeed grasps the essential features of our data (Figs. 3 and 4),
if we use $\xi_{\mathrm{ve}}$ and $\tau_{\mathrm{ve}}$ as adjustable
parameters. In addition, to obtain $\tau_{q}=1/Dq^{2}$, we need to use the
apparent mesoscopic viscosity $\eta_{\mathrm{app}}$ in Eq. (\ref{Eq2Diff}). The
predictions for the two uncoupled modes (pure diffusion and pure viscoelastic
relaxation) are indicated by the dashed curves in Fig.3. While the diffusion
relaxation time changes rapidly when the critical point is approached, the
viscoelastic relaxation time exhibits a regular dependence on temperature.
While the original uncoupled modes would cross each other at a certain
temperature, the coupling produces two effective modes that ''avoid crossing''
each other very much similar to the well known avoided crossing of two 
coupled energies~\cite{Landau77}. 
Therefore, the microrheological characteristics can be deduced from
scattering data in a near-critical fluid, since one can vary the 
diffusion relaxation time over many orders, thus probing the relevant 
viscoelastic times over the same range. 

While $\xi_{\mathrm{ve}}$ (as expected~\cite{Doi92,Onuki02}) appears to be
proportional to the viscosity, it was not clear a priori which viscosity is
the appropriate quantity, the mesoscopic $\eta_{\mathrm{app}}$ or the
macroscopic $\eta$ at zero shear rate. Our study has shown that $\xi
_{\mathrm{ve}}$ is proportional to the apparent (''mesoscopic'') viscosity
measured by DLS. A further notable point is the shift in Fig.3 between the
calculated diffusion mode (long-dashed curve) and the observed slow mode
($\tau_{-}$, solid curve). The data suggest that $\tau_{-}$ is slowed down at
nanoscales by a factor $q^{2}\xi_{\mathrm{ve}}^{2}$ with respect to the
diffusion mode. For example, at $M=11.4$ million and a scattering angle of $30%
{{}^\circ}$, where length scales of about $q^{-1}=137$ nm are probed and where
$\xi_{\mathrm{ve}}$ reaches 200 nm, the slow mode is shifted from 0.4 s to 1.5
s. We may attribute this anomalous slowing down of the fluctuations to
''diffusion trapped by viscoelasticity at the nanoscale'' and we expect this
effect to be ubiquitous in viscoelastic systems. Note that this effect of
additional slowing down at smaller scales (large $q$) is opposite to the
famous critical slowing down, which becomes more pronounced at larger scales
(small $q$). In Fig.4 the experimental amplitudes of these effective dynamic
modes are compared with the theoretical ones calculated with Eq. (\ref{Eq5f}).
We submit that our interpretation of the coupled modes on the basis of the
Brochard-De Gennes theory does account for the essential physics of the
phenomenon. The analysis of the observed avoided-crossing of two coupled 
modes has a good sensitivity because the amplitudes of the two effective 
modes become of the same order of magnitude in the avoided-crossing 
domain (Fig.4).

The key results obtained in our study are significant far beyond just the
near-critical polymer solutions investigated. The coupling between
diffusion-like and structure relaxation modes is expected whenever such modes
are close to each other and thus scanning the diffusivity decay times by any
means (varying composition, temperature, or pressure) will reveal the
structural relaxation. Our results are relevant for a variety of complex
fluids in which critical phenomena couple with a mesoscopic structure and/or
with viscoelastic relaxation. These include polymers in supercritical
fluids~\cite{Melnichenko00}, polymer blends~\cite{Frielinghaus01}, polymer
solutions under shear~\cite{Dixon92}, and
microemulsions~\cite{Rouch93,Hellweg00}, as well as systems important in the
life sciences, such as solutions of polyelectrolytes or
biopolymers~\cite{Nishida01,Chirico96}.

We conclude by highlighting the possible practical applications of studying
the coupling between diffusive relaxation of critical fluctuations and
structural relaxation. Because of this coupling, dynamic light scattering of
critical fluctuations becomes a new tool for measuring the rheological
properties of near-critical complex fluids. That is, by performing
non-invasive DLS measurements on a sample, we can obtain quantitative
information concerning its microrheological properties. Our proposed approach
may be termed ''critical microrheology'' and does not require the addition of
probe particles to the fluid. The uniqueness of critical microrheology is its
ability to scan diffusive decay time of fluctuations at a given length scale
$q^{-1}$ over several orders of magnitude, and thereby probe viscoelastic
relaxation as an intrinsic fluid property. Moreover, instances have been
reported where microrheological measurements are inconsistent with macroscopic
rheology~\cite{VanZanten00}. ''Critical microrheology'' experiments may
clarify the nature and extent of these discrepancies. By selecting appropriate
solvents for bringing systems into a near-critical state one should be able to
probe structural relaxation of a variety of macromolecular species in solutions.

We acknowledge some valuable discussion with M. R. Moldover.
The research was supported by the Office of Basic Energy Sciences of the US
Department of Energy under Grant No. DE-FG-02-95ER-14509.

\bigskip

\section{Figures captions}

Figure 1. Apparent mesoscopic viscosity of a solution of polystyrene
($M=195,900$) in cyclohexane as a function of 
$\varepsilon=(T-T_{\mathrm{c}})/T$ obtained by fitting the
experimental light-scattering data to the mode-coupling theory. The dotted
curve represents the viscosity of the solvent (cyclohexane) and the dashed
curve represents the macroscopic viscosity of the same solution \cite{Lao75}.

Figure 2. Mesoscopic diffusion coefficient of a solution of polystyrene
($M=195,900$)in cyclohexane as a function of $\varepsilon=(T-T_{\mathrm{c}})/T$ 
measured at three scattering angles.
The symbols represent experimental data, while the curves represent the
critical contribution predicted by mode-coupling theory with the
mesoscopic viscosity shown in Fig.1.

Figure 3. Dynamic modes in a solution of polystyrene ($M=11.4$ million) in
cyclohexane for $q$, corresponding to a scattering angle of $30{^\circ}$. 
Open symbols represent the experimental relaxation times of the two
observed modes. The solid curves represent the relaxation times of the
effective ''slow'' and ''fast'' modes, calculated with Eq. (\ref{Eq4tauf}). The
long-dashed curve represents the uncoupled critical-diffusion decay time. The
short-dashed curve represents the uncoupled viscoelastic relaxation time.

Figure 4. Amplitudes of the two effective dynamic modes as a function of 
$\varepsilon=(T-T_{\mathrm{c}})/T$ in a near-critical polymer solution. 
Solid curves are theoretical predictions for the amplitudes (Eq. (\ref{Eq5f})).

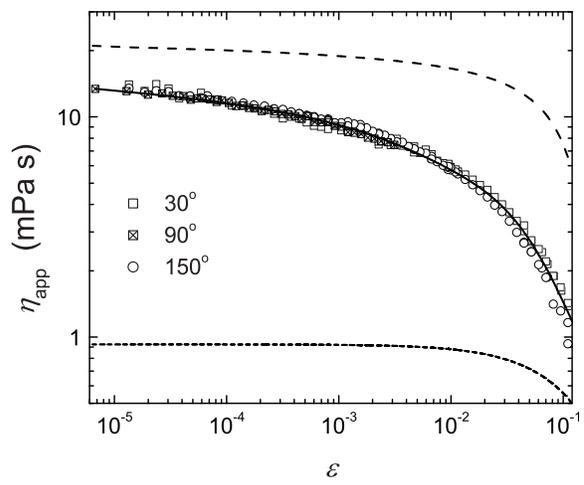
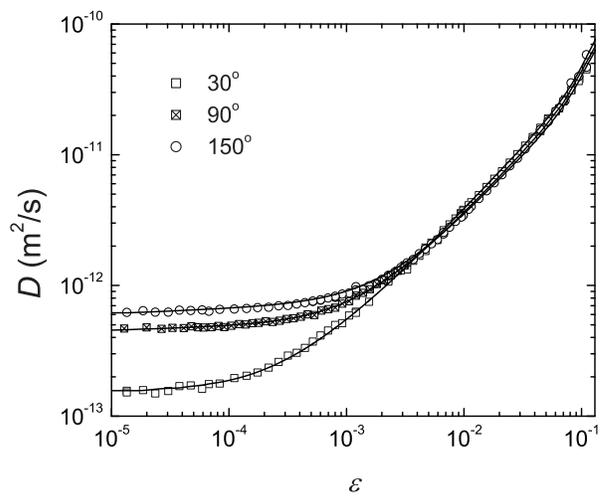
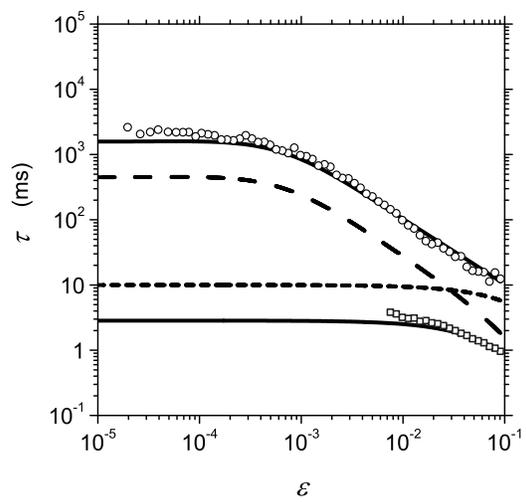
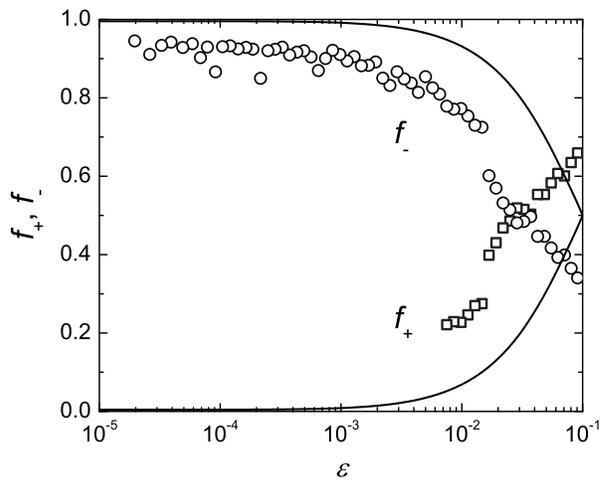
\end{document}